\documentstyle[12pt]{article}

\begin{document}

\def\unit{\hbox to 3.3pt{\hskip1.3pt \vrule height 7pt width .4pt \hskip.7pt
\vrule height 7.85pt width .4pt \kern-2.4pt
\hrulefill \kern-3pt
\raise 4pt\hbox{\char'40}}}
\def\II{{\unit}}
\def\cM {{\cal{M}}}
\def\half{{\textstyle {1 \over 2}}}
%       LOCAL DEFINITIONS

\newcommand{\mathbold}[1]{\mbox{\rm\bf #1}}
\newcommand{\mrm}[1]{\mbox{\rm #1}}
\newcommand{\gtrsim}
{\ \rlap{\raise 2pt\hbox{$>$}}{\lower 2pt \hbox{$\sim$}}\ }
\newcommand{\lessim}
{\ \rlap{\raise 2pt\hbox{$<$}}{\lower 2pt \hbox{$\sim$}}\ }

\newcommand{\ea}{{ et al.}}
\newcommand{\ib}{{\it ibid.\ }}

\def\calo{{\cal O}}
\def\call{{\cal L}}
\def\calm{{\cal M}}
\def\ie{{\it i.e.}}
\def\eg{{\it e.g.}}
\def\msusy{m_{\rm S}}
\def\be{\begin{equation}}
\def\ee{\end{equation}}
\def\bea{\begin{eqnarray}}
\def\eea{\end{eqnarray}}
\def\identity{1 \hspace{-.085cm}{\rm l}}
\def\char{\tilde\chi^+}
\def\neu{\tilde\chi^0}
\def\neub{\tilde\chi^{0}{'}}

\begin{titlepage}
\begin{center}
\hfill CERN--TH/95--337
\end{center}
\vskip 1in
\center{\Large\bf  Mass Degeneracy of the Higgsinos}
 \vskip .6in
\center{{\sc Gian F. Giudice\footnote[1]{On leave of absence from
 INFN Sezione di Padova, Padua, Italy.} and Alex
Pomarol}}
\center{{\it Theory Division, CERN}\\{\it CH-1211 Geneva 23,
Switzerland}}
\vskip .6in
\begin{abstract}
The search for  charginos   and neutralinos  at LEP2
can become problematic if these particles are almost mass degenerate
with
 the lightest neutralino.
Unfortunately this is the case  in the region where these particles
are higgsino-like.
 We show  that, in this region,
 radiative  corrections to
the higgsino mass splittings
 can be as large as the tree-level
values, if the mixing between the two stop states is large.
We also show that the degree of degeneracy of the higgsinos
 substantially increases
 if a large phase is present in the
higgsino mass term $\mu$.

\vfill

\noindent CERN--TH/95--337\hfill\\
\noindent December 1995\hfill
\end{abstract}
\end{titlepage}

\newpage

\baselineskip20pt
\pagestyle{plain}

%%%%%%%%%%%%%%%%%%%%%%%%%%%%%%%%%%%%%%%%%%%%%%%%%%%%%%%%%%%%%%%%%%%%%%

\makeatletter
\setlength{\clubpenalty}{10000}
\setlength{\widowpenalty}{10000}
\setlength{\displaywidowpenalty}{10000}

\vbadness = 5000
\hbadness = 5000
\tolerance= 1000

\arraycolsep 2pt

\footnotesep 14pt

\if@twoside
\oddsidemargin -17pt \evensidemargin 00pt \marginparwidth 85pt
\else \oddsidemargin 00pt \evensidemargin 00pt
\fi
\topmargin 00pt \headheight 00pt \headsep 00pt
\footheight 12pt \footskip 30pt
\textheight 232mm \textwidth 160mm

\let\@eqnsel = \hfil

\expandafter\ifx\csname mathrm\endcsname\relax\def\mathrm#1{{\rm #1}}\fi
\@ifundefined{mathrm}{\def\mathrm#1{{\rm #1}}}{\relax}

\makeatother

\unitlength1cm
\baselineskip22pt

%%%%%%%%%%%%%%%%%%%%%%%%%%%%%%%%%%%%%%%%%%%%%%%%%%%%%%%%%

The search for charginos ($\char$) at LEP2 is one of the most
 promising ways of discovering low-energy supersymmetry.
 If the $\char$ decays into the lightest neutralino ($\neu$) and a virtual
$W^+$, it can be discovered at LEP2 (with a $\int\call=500$ pb$^{-1}$)
whenever its production cross section is larger than about
 $0.1$--$0.3$ pb  and  $m_{\neu}$ is within the range $m_{\neu}\gtrsim 20$ GeV
and $m_{\char}- m_{\neu}\gtrsim 5$--$10$ GeV \cite{lepr}.
Therefore, the chargino can be discovered almost up to the LEP2 kinematical
 limit, unless one of the following three conditions occurs:

{\it i}) The sneutrino ($\tilde\nu$) is light
 and the chargino is mainly gaugino-like. In this case the $\tilde\nu$
 $t$-channel exchange
interferes destructively with the gauge-boson exchange and can
 reduce the chargino production cross section
below the minimum values required for observability, $0.1$--$0.3$ pb.
However, in a large fraction
 of the parameter space where this effect is important, the two-body decay
 mode $\char\rightarrow\tilde\nu l^+$ is kinematically allowed and dominates
 over the conventional three-body decays  $\char\rightarrow\neu l^+\nu$,
 $\neu \bar qq$. The resulting signal,
quite similar to the one caused by slepton pair production, allows the
chargino search for much smaller production cross section, possibly
as small as $20$--$60$ fb \cite{lepr}.

{\it ii}) The $\neu$ is very light ($m_{\neu}\lessim 20$ GeV). In this case
 the $\char$-detection
 efficiency diminishes, as the decrease in
missing invariant mass makes the signal more
similar to  the $W^+W^-$ background.
Such a light $\neu$
 is allowed by LEP1 data, if the weak-gaugino mass ($M$)
is small\footnote{The parameter region
where $M$ and the higgsino mass ($\mu$) are both small has been recently
studied in ref.~\cite{fpt}.}, $M\ll M_W$.
It is however ruled out by gluino ($\tilde g$) searches at the
Tevatron \cite{tev} in models which assume gaugino mass unification,
$ M=\alpha M_{\tilde g}/(\alpha_s\sin^2\theta_W)$. It should
also be mentioned that no study has attempted to optimize the analysis
in the low-$m_{\neu}$ region, while specially designed experimental
cuts could improve the chargino  detection efficiency.

{\it iii}) $\Delta_+\equiv m_{\char}-m_{\neu}$ is small ($\Delta_+\lessim
 5$--$10$ GeV) and the $\char$ detection is problematic because of the lack
of energy of the visible decay products \cite{lepr}.

In this letter, we concentrate on case ({\it iii}).
 Small values of $\Delta_+$ occur
when the gaugino masses are much larger than $M_W$.
In this region, the lightest chargino and the two lightest
neutralinos ($\neu$,$\neub$)
are mainly higgsino-like and are nearly degenerate\footnote{It
 has been recently shown \cite{cdg} that
single-photon tagging cannot be used to observe charginos in the higgsino
region under consideration. Ref.~\cite{cdg} has also considered a gaugino-like
region where $\Delta_+$ can become small in models without
gaugino mass unification.} with mass $\sim\mu$.
In ref.~\cite{acmw} it has been suggested
 that this parameter region
 ($M\gg M_W$ and $\mu\sim M_W$), although problematic for
 chargino searches,
can nevertheless be covered by neutralino searches. Indeed, in this case the
$\neu$--$\neub$ production cross section
at LEP2 is large ($\sim$ pb) and
 the mass difference $\Delta_0\equiv m_{\neub}-
m_{\neu}$
is  always greater than $\Delta_+$,
allowing a better identification of the visible decay products
than in the chargino case.
It is also worth recalling that the parameter region
where $M\gg M_W$ and $\mu\sim M_W$ has a special interest as a
light higgsino-like chargino (together with a light stop) can increase the
 Standard Model prediction for $R_b$ \cite{rb}.

Here
 we point out that, for $M\gg M_W$ and $\mu\sim M_W$,
 $\Delta_0$ and $\Delta_+$
receive  one-loop corrections
 proportional to $m_t^3$, which
can be as large as their tree-level values.
We will also show that in the small $\tan\beta$ region, the tree-level values
of $\Delta_0$ and $\Delta_+$ can be reduced if the $\mu$ parameter
has a non-trivial phase.
Finally, we will comment on the implications for chargino and neutralino
searches at LEP2.

First let us recall that, in the higgsino region under consideration
($M\gg M_W$, $\mu\sim M_W$), the tree-level values
of $\Delta_0$ and $\Delta_+$ are well approximated by
a $1/M$ expansion\footnote{
Here we take $M$ to be real and positive and $\mu$ to be real
following the sign conventions of ref.~\cite{hk}.
 A complex $\mu$ will be considered
 later.}
\bea
\Delta_0&=&2a\frac{M^2_W}{M}+2b\sin 2\beta\,\frac{\mu M^2_W}{M^2}+\calo(1/M^3)
\, ,\nonumber\\
\Delta_+&=&\big[a+(a-1)\varepsilon\sin 2\beta\big]\frac{M^2_W}{M}
+\big[(b-1)+b\varepsilon\sin 2\beta\big]\frac{|\mu| M^2_W}{M^2}\nonumber\\
&-&\frac{a^2}{2}\cos^22\beta\frac{M^4_W}{|\mu| M^2}+\calo(1/M^3)\, ,
\label{delt}
\eea
where
\bea
a&\equiv& \frac{4}{5}+\frac{1}{2}\left(\frac{M}{M{'}}
\tan^2\theta_W-\frac{3}{5}\right)\, ,\nonumber\\
b&\equiv& \frac{1}{2}\left(1+\frac{9}{25\tan^2\theta_W}\right)+
\frac{1}{2}\left(\frac{M}{M{'}}\tan^2\theta_W-\frac{3}{5}\right)
\left(\frac{M}{M{'}}+\frac{3}{5\tan^2\theta_W}\right)\, .
\eea
Here $\tan\beta$ is the ratio of the two Higgs vacuum expectation values,
$\varepsilon=\mu/|\mu|$ and $M{'}$ is the hypercharge gaugino mass.
 The  expansions in eq.~(\ref{delt})
break down when $\mu\rightarrow 0$ but this is
 not relevant here since, for $M\gg M_W$, LEP1 data  require $\mu>M_Z/2$.

By comparing the leading $1/M$ terms in eq.~(\ref{delt}), one finds that
$\Delta_0>\Delta_+>0$ for any (positive) value of $M{'}$.
Assuming the gaugino mass unification condition
 $M{'}=\frac{5}{3}\tan^2\theta_W M$, we obtain:
\be
\Delta_+=\frac{1}{2}\left(1-\frac{\varepsilon\sin 2\beta}{4}
\right)\Delta_0+\calo(1/M^2)\, .
\label{cora}
\ee
Notice also that the critical region
for chargino searches ($\Delta_+\lessim 10$ GeV) occurs when $M\gtrsim
 400$--$600$ GeV, depending on $\tan\beta$.
For the neutralinos, however, we have
 $\Delta_0\lessim 10$ GeV  for  $M\gtrsim 1$ TeV.

General expressions for the one-loop corrections to chargino
and neutralino masses are given in ref.~\cite{loop}.
To obtain simple analytical formulae we have computed the radiative
corrections in the limit $M\rightarrow\infty$.
The only contributions arise from heavy quark--squark loops
 and $\gamma(Z)$--higgsino
loops\footnote{The overall sign of $\delta\Delta_0$
 is chosen by assuming that the lightest
neutralino is determined by the tree-level relation in eq.~(\ref{delt}).}:
\bea
\delta\Delta_0&=&
2G_t^2m_t\sin 2\theta_t\sum_{i=1,2}(-1)^{i+1}B_0(\mu^2,m^2_t,m^2_{\tilde t_i})
+(t\leftrightarrow b)\, ,\label{a}\\
\delta\Delta_+&=&
\frac{\delta\Delta_0}{2}-
\sum_{i=1,2}\bigg[G_tG_bm_t\sin 2\theta_b(-1)^{i+1}B_0(\mu^2,m^2_t,
m^2_{\tilde b_i})\nonumber\\
&-&
|\mu|G^2_tB_1(\mu^2,m^2_t,m^2_{\tilde t_i})+|\mu|H^{tb}_i
B_1(\mu^2,m^2_t,m^2_{\tilde b_i})
+(t\leftrightarrow b)\bigg]\label{b}\\
&+&\frac{\alpha}{\pi}|\mu|\left[B_0(\mu^2,\mu^2,0)-
B_0(\mu^2,\mu^2,M^2_Z)-\frac{1}{2}
B_1(\mu^2,\mu^2,0)+\frac{1}{2}B_1(\mu^2,\mu^2,M^2_Z)\right]\, ,
\nonumber
\eea
where
\bea
H^{tb}&\equiv&(G^2_t\cos^2\theta_b+G^2_b\sin^2\theta_b,
G^2_t\sin^2\theta_b+G^2_b\cos^2\theta_b)\, ,\nonumber\\
G_t&\equiv&\sqrt{\frac{3G_F}{8\sqrt{2}\pi^2}}\,\frac{m_t}{\sin\beta}\, ,
\ \ \ \
G_b\equiv\sqrt{\frac{3G_F}{8\sqrt{2}\pi^2}}\,\frac{m_b}{\cos\beta}\, .
\eea
Explicit expressions for the functions $B_0$ and $B_1$, defined as
\be
B_n(p^2,m_1^2,m_2^2)=-\int^1_0 dx x^n\log[-p^2x(1-x)+m^2_1(1-x)
+m^2_2x-i\epsilon]\ \ \ (n=0,1)\, ,
\ee
can be found in ref.~\cite{velt}. The stop mixing angle $\theta_t$
is defined such that $\tilde t_1=\cos\theta_t\tilde t_L+
\sin\theta_t\tilde t_R$ is the heavier mass eigenstate and
$\tilde t_2=-\sin\theta_t\tilde t_L+
\cos\theta_t\tilde t_R$ is the lighter one; the sbottom mixing angle
$\theta_b$ is defined analogously.

As expected, all terms in eqs.~(\ref{a}) and (\ref{b}) vanish in the limit
of exact electroweak symmetry.
In this limit, all higgsino mass terms other than $\mu$ are forbidden.
Notice that eq.~(\ref{a}) also vanishes when the stop mixing angle
is zero (neglecting the term proportional to $m_b$),
in spite of the presence of the electroweak breaking top-quark mass $m_t$.
This happens because, in order to generate a non-vanishing
$\delta\Delta_{0}$, one has to break an $R$-symmetry under which the
Higgs superfields $H_1$ and $H_2$ and the top quark superfields
$Q_L$, $\bar U_R$  carry charges $R=\{2,0,0,2\}$, respectively.
  This $R$-symmetry, although not broken by the top mass,
 is violated by stop left--right mixing terms.

The dominant contribution in eq.~(\ref{a})
comes from the top--stop loops and it is approximately given by
\be
\delta\Delta_0\simeq 2G^2_tm_t\sin 2\theta_t\log\left[
\frac{{\rm max}(m^2_t,m^2_{\tilde t_2})}{m^2_{\tilde t_1}}\right]\, ,
\label{c}
\ee
for large mass splitting
$(m^2_{\tilde t_1}\gg m^2_{\tilde t_2})$.
For
maximal stop left--right mixing
 ($\sin 2\theta_t\simeq 1$)  eq.~(\ref{c}) predicts
$|\delta\Delta_0|\sim 12$ GeV$\times \left(\frac{m_t}{180{\rm GeV}}\right)^3
\left(\frac{1}{\sin^2\beta}\right)$
 for a  stop mass
 $m_{\tilde t_1}\sim 1$ TeV.
Under these conditions, the one-loop corrections to $\Delta_0$ can easily be
 of the order of the tree-level value.
Notice, however, that these conditions on the stop mass parameters
are not the ones that
 maximize the supersymmetric corrections to $R_b$ \cite{rb}.
The sign of the corrections
depends on the sign of $\sin 2\theta_t$, which in turn
 is proportional to the unknown
value of the stop left--right mixing. Therefore $\delta\Delta_{0}$ can
either enhance or suppress the tree-level result.

The contributions proportional to $|\mu|$ in eq.~(\ref{b})
never amount to more than a few GeV for any value of $|\mu|$ relevant
to LEP2 searches.
Therefore when $\tan\beta$ is not too large, eq.~(\ref{b}) approximately
 predicts:
\be
\delta\Delta_+\simeq \frac{\delta\Delta_0}{2}\label{cor}\, ,
\ee
which mimics  the tree-level relation of eq.~(\ref{cora}).
 On the other hand,
for large $\tan\beta$, the second term in eq.~(\ref{b}) can become
 important  and may destroy the correlation between $\delta\Delta_0$
and $\delta\Delta_+$ given by
eq.~(\ref{cor}).

Since the loop corrections to $\Delta_0$ and $\Delta_+$
are coming from electroweak breaking effects, it is necessary to check
whether the same choice of parameters also leads to unacceptably
large corrections
to the electroweak observables at LEP1.
We have verified that there are  no large effects for either $\epsilon_2$
or $\epsilon_3$ \cite{ab} (or equivalently $U$ or $S$ \cite{pes}).
For instance, taking maximal mass splittings
 $m^2_{\tilde t_1}\gg m^2_{\tilde t_2}>M^2_Z$ and
 $m^2_{\tilde b_1}\gg m^2_{\tilde b_2}>M^2_Z$,
 we find \cite{dhy}
\bea
S&=&\frac{4\sin^2\theta_W}{\alpha}\epsilon_3=\frac{1}{12\pi}\bigg[
\log\frac{m^2_{\tilde b_1}}{m^2_{\tilde t_1}}+\sin^2\theta_t
(4-3\sin^2\theta_t)\log\frac{m^2_{\tilde t_1}}{m^2_{\tilde t_2}}\nonumber\\
&+&\sin^2\theta_b(2-3\sin^2\theta_b)
\log\frac{m^2_{\tilde b_1}}{m^2_{\tilde b_2}}
-\frac{5}{4}(\sin^22\theta_t+\sin^22\theta_b)\bigg]\, ,
\eea
which is smaller than about $0.1$, even for maximal squark left--right mixing
and mass splittings as large as
 $m_{\tilde t_1}/ m_{\tilde t_2}\sim
 m_{\tilde b_1}/m_{\tilde b_2}\sim 10$.

More important is the constraint coming from the $\rho$ parameter.
The contribution from  stop and sbottom loops gives \cite{rho}
\bea
\Delta\rho&=&\frac{3G_F}{4\sqrt{2}\pi^2}\Bigg\{
\cos^2\theta_t\Big[\cos^2\theta_bf(m^2_{\tilde t_1},m^2_{\tilde b_1})
+\sin^2\theta_bf(m^2_{\tilde t_1},m^2_{\tilde b_2})\Big]\nonumber\\
&+&
\sin^2\theta_t\Big[\cos^2\theta_bf(m^2_{\tilde t_2},m^2_{\tilde b_1})
+\sin^2\theta_bf(m^2_{\tilde t_2},m^2_{\tilde b_2})\Big]\nonumber\\
&-&
\cos^2\theta_t\sin^2\theta_tf(m^2_{\tilde t_1},m^2_{\tilde t_2})-
\cos^2\theta_b\sin^2\theta_bf(m^2_{\tilde b_1},m^2_{\tilde b_2})\Bigg\}
\, ,\label{rh}
\eea
where
\be
f(x,y)=\frac{xy}{x-y}\log\frac{y}{x}+\frac{x+y}{2}\, .
\ee
In order to study the predictions on $\delta\Delta_+$
and  $\delta\Delta_0$ compatible with the present constraint on the $\rho$
parameter,
we first define the stop and sbottom squark mass matrices as
\bea
m^2_{\tilde t}&=&\pmatrix{
{ m^2_{\widetilde Q_L}}+m^2_t+
(\frac{1}{2}-\frac{2}{3}\sin^2
\theta_W)\cos 2\beta M_Z^2
          & m_t( A_t-\mu\cot\beta)\cr
m_t( A_t-\mu\cot\beta)&
  m^2_{\tilde t_R}+m^2_t+\frac{2}{3}
\sin^2\theta_W\cos 2\beta M_Z^2  }\, ,\nonumber\\
\label{squarkmass}\\
m^2_{\tilde b}&=&\pmatrix{
{ m^2_{\widetilde Q_L}}+m^2_b-
(\frac{1}{2}-\frac{1}{3}\sin^2
\theta_W)\cos 2\beta M_Z^2          & m_b( A_b-\mu\tan\beta)\cr
m_b( A_b-\mu\tan\beta)&
  m^2_{\tilde b_R}+m^2_b-\frac{1}{3}
\sin^2\theta_W\cos 2\beta M_Z^2  }\, .\nonumber
\eea
Since we do not want to rely on specific model-dependent assumptions, we will
 treat the supersymmetry-breaking parameters $m^2_{\widetilde Q_L}$,
$m^2_{\tilde t_R}$, $m^2_{\tilde b_R}$, ${ A_t}$ and ${ A_b}$
 as free variables.
The result of varying these five
 supersymmetry-breaking
 parameters (and the sign of $\mu$) compatibly with
the constraints $\Delta\rho<1\times 10^{-3}$ or $3\times 10^{-3}$ is shown in
fig.~1.
We have chosen $m_t=180$ GeV and $|\mu|=80$ GeV, but the dependence on $|\mu|$
is insignificant within the range of interest for LEP2.
 Figure~1a corresponds to the case of minimal
$\tan\beta$ consistent with perturbativity up to the GUT scale,
$\sin\beta\simeq m_t/(195$ GeV). Figure~1b corresponds
to the case of maximal $\tan\beta$, $\tan\beta\simeq m_t/m_b$.
The stop left--right mixing parameter $A_t$
plays a crucial role, since $\delta\Delta_0$ tends to zero in the limit
$\sin 2\theta_t\rightarrow 0$. The largest effects on $\delta\Delta_0$
are obtained for maximal stop mixing and mass splitting,
 $m^2_{\widetilde Q_L}\sim m^2_{\tilde t_R}\sim A_tm_t$.
 Figure~1 shows
the comparison between the region of  $\delta\Delta_+ - \delta\Delta_0$
values which can be obtained by requiring $|\bar A_t|
\equiv |2A_t/(m_{\widetilde Q_L}+m_{\tilde t_R})|<1$ and that where
$|\bar A_t|<3$. For $|\bar A_t|<1$, the regions
in the    $\delta\Delta_+ - \delta\Delta_0$ space where
$\Delta\rho<1\times 10^{-3}$ and
 $\Delta\rho<3\times 10^{-3}$ are about the same.

Although constraints from the $\rho$ parameter reduce the maximum
values of $\delta\Delta_0$ with respect to
 the estimate  in eq.~(\ref{c}), the effect of the one-loop corrections
to $\Delta_0$
can still be sizeable, possibly of the order of the tree-level contributions.
The relation between
 $\delta\Delta_0$ and $\delta\Delta_+$ in eq.~(\ref{cor}) is a good
approximation for the small $\tan\beta$ (see fig.~1a), but
deviations can appear for very large values of $\tan\beta$
(see fig.~1b).

Finally, we want to show how
the values of
$\Delta_+$ and  $\Delta_0$  in eq.~(\ref{delt}) can  be modified when
 the $\mu$ parameter is allowed to be complex.
The  electric dipole moments of the neutron and electron
put severe constraints on the phase of $\mu$ \cite{edm}.
Nevertheless these constraints can be relaxed if the masses of the
first generation of squarks and sleptons
 or the gaugino masses are much larger than $\mu$.
For example, if the first generation of squarks and sleptons masses
 are larger than $\sim 1$ TeV,
 no limits can be placed on
the phase of $\mu$.
Such large masses are allowed in scenarios with non-universal
soft masses without   problems of fine tuning \cite{us}.
For a complex $\mu$ parameter, $\mu=|\mu|e^{i\varphi}$,
and $M$ real and positive, we find:
\bea
\Delta_0&=&\Bigg|2ra\frac{M^2_W}{M}+\frac{2\sin 2\beta\cos\varphi}{r|\mu|}
\bigg[b|\mu|^2\frac{M^2_W}{M^2}
+(r^2-1)a^2\frac{M^4_W}{M^2}\bigg]\Bigg|+\calo (1/M^3)\, ,\nonumber\\
\Delta_+&=&\frac{\Delta_0}{2}+\sin2\beta\cos\varphi(a-1)\frac{M^2_W}{M}+
|\mu|(b-1)\frac{M^2_W}{M^2}\nonumber\\
&+&\bigg[1-r^2+(2r^2-2-\cos^2 2\beta)a^2\bigg]
\frac{M^4_W}{2|\mu|M^2}+\calo (1/M^3)\, ,
\label{deltm}
\eea
where
\be
r=\sqrt{1-\sin^2 2\beta\sin^2\varphi}\, .
\ee
By comparing the leading $1/M$ terms in eq.~(\ref{deltm}), we find
$\Delta_0>\Delta_+>0$, for any value of $\varphi$ and $M{'}/M$.
If we assume the unification condition $M{'}/M=5\tan^2\theta_W/3$,
then $\Delta_+/\Delta_0<5/8$ for any $\varphi$.
Notice that the effect of a non-trivial
phase is always to reduce the value of $\Delta_0$ with respect to
the cases $\varphi=0,\pi$.
However, for large $\tan\beta$,  the dependence
of $\Delta_+$ and  $\Delta_0$ on the phase $\varphi$ disappears.
This happens because, in the limit
 $\langle  H_1\rangle\rightarrow 0$, one can  rotate
away the phase $\varphi$ from the chargino and neutralino mass matrices.
On the contrary, for small  $\tan\beta$,
the effect of $\varphi$ on $\Delta_+$ and  $\Delta_0$
can be significant.
In the limit  $\tan\beta\rightarrow 1$ and
 $\varphi\rightarrow\pi/2$,
 the leading $1/M$
contribution to eq.~(\ref{deltm}) vanishes
and therefore
 $\Delta_+$ and  $\Delta_0$  are drastically reduced.
In this case,  $\Delta_+$ can even become negative
and the chargino may be the lightest supersymmetric particle.

In conclusion, we have shown that in the higgsino region ($M\gg M_W$, $\mu\sim
M_W$) the one-loop corrections to the mass splittings $\Delta_+$
and $\Delta_0$ can be of the same order of the tree-level
values. The effect is significant only if the mixing angle and the
mass splitting of the two stop states are large.
Depending on the sign of the stop mixing angle,  these corrections can
decrease or increase the mass splittings
$\Delta_+$
and $\Delta_0$. Thus, radiative corrections can make charginos and neutralinos
very degenerate in mass for values of $M$ much smaller than
previously thought. The opposite can also happen,
 and radiative corrections can
allow the discovery of
  charginos and neutralinos at LEP2 in parameter
regions predicted to be problematic by the tree-level relations.
The values of $\Delta_+$
and $\Delta_0$
 can also be reduced
if $\mu$ has a  large phase and $\tan\beta\sim 1$.

If $\tan\beta$ is not too large,
the one-loop corrections to $\Delta_+$
and $\Delta_0$
 are correlated, see eq.~(\ref{cor}) and fig.~1a, and mimic the tree-level
relation, see eq.~(\ref{cora}). Therefore
the conclusion of ref.~\cite{acmw}
that neutralino search is an important experimental tool in the
study of the higgsino region remains valid.
However the
relation between $\Delta_+$, $\Delta_0$
and the gaugino masses can be lost.
Finally, if $\tan\beta$ is extremely large,
 sbottom loop corrections can
 partially spoil the $\Delta_+ - \Delta_0$ correlation,
 and higgsino mass splittings
more critically depend on the parameter choice.
\vskip .5in
We thank M. Carena, M. Mangano and C. Wagner
for interesting discussions and very helpful suggestions.
We also thank D. Pierce for useful correspondence.

\newpage
\noindent{\Large{\bf Figure Captions}}
\vskip .2in
\noindent{\bf Fig.~1:} {Region of values in the
 $\delta\Delta_+ - \delta\Delta_0$ plane
 obtained by varying the supersymmetry-breaking
parameters defined in eq.~(\ref{squarkmass}), with
 $m_t=180$ GeV, $|\mu|=80$ GeV, $\tan\beta=2.4$ (fig.~1a)
and $\tan\beta\simeq m_t/m_b$ (fig.~1b). The different lines
correspond to different constraints on $|\bar A_{t,b}|$ and $\Delta\rho$.}
\pagebreak


\newpage

\begin{thebibliography}{99}

\frenchspacing
\bibitem{lepr}
G. Altarelli et al., {\it Interim Report on Physics Motivations
for an Energy Upgrade of LEP2}, CERN-TH/95--151, CERN-PPE/95--78;
G. Altarelli et al., {\it Report of the Workshop on Physics at LEP2}, in
preparation.
\bibitem{fpt}
J.L. Feng, N. Polonsky and  S. Thomas, preprint  SLAC-PUB-95-7050 (1995).
\bibitem{tev}
D.R. Claes for the D0 Coll., {\it Proceedings of the
10th Topical Workshop on Proton-Antiproton Collider}, Batavia, IL (USA),
9--13 May 1995; L. Nodulman for the CDF Coll., {\it
Proceedings of the International
Europhysics Conference on High Energy Physics} (HEP 95), Brussels,
Belgium, 27 July--2 August 1995.
\bibitem{cdg}
C.H. Chen, M. Drees and  J. Gunion, preprint UCD-95-43 (1995).
\bibitem{acmw}
S. Ambrosanio, B. Mele, M. Carena and  C.E.M. Wagner, preprint
CERN-TH/95-286 (1995).
\bibitem{rb}
G. Altarelli, R. Barbieri and F. Caravaglios, Phys. Lett. {\bf B314}
(1993) 357;
 J.D. Wells, C. Kolda and G.L. Kane, Phys. Lett.
{\bf B338} (1994) 219;
 D. Garcia, A. Jimenez and J. Sola,
Phys. Lett. {\bf B347} (1995) 309, 321; E {\bf B351} (1994) 278.
\bibitem{hk}
H.E. Haber and G.L. Kane, Phys. Rep. {\bf 117} (1985) 75.
\bibitem{loop}
D. Pierce  and
 A. Papadopoulos, Phys. Rev. {\bf D50} (1994) 565;
Nucl. Phys. {\bf B430} (1994) 278;
A.B. Lahanas, K. Tamvakis and  N.D. Tracas,
Phys. Lett. {\bf B324} (1994) 387.
\bibitem{velt}
G. Passarino and  M. Veltman,  Nucl. Phys. {\bf B160} (1979) 151.
\bibitem{ab}
G. Altarelli  and  R. Barbieri,  Phys. Lett. {\bf B253} (1991) 161.
\bibitem{pes}
M.E. Peskin and T. Takeuchi, Phys. Rev. Lett. {\bf 65} (1990) 964.
\bibitem{dhy}
M. Drees, K. Hagiwara and A. Yamada,  Phys. Rev. {\bf D45} (1992) 1725.
\bibitem{rho}
R. Barbieri and L. Maiani, Nucl. Phys. {\bf B224} (1983) 32;
C.S. Lim, T. Inami and N. Sakai, Phys. Rev. {\bf D29} (1984) 1488;
 Z. Hioki, Prog. Theor. Phys. {\bf 73} (1985) 1283;
J.A. Grifols and J. Sola, Nucl. Phys. {\bf B253} (1985) 47.
\bibitem{edm}
W. Buchm\"uller and D. Wyler, Phys. Lett. {\bf B121} (1983) 321;
J. Polchinski and M.B. Wise, Phys. Lett. {\bf B125} (1983) 393.
\bibitem{us}
S. Dimopoulos and G.F. Giudice, Phys. Lett. {\bf B357} (1995) 573;
A. Pomarol and D. Tommasini, preprint CERN--TH/95--207 (1995) (hep-ph/9507462).
\end{thebibliography}
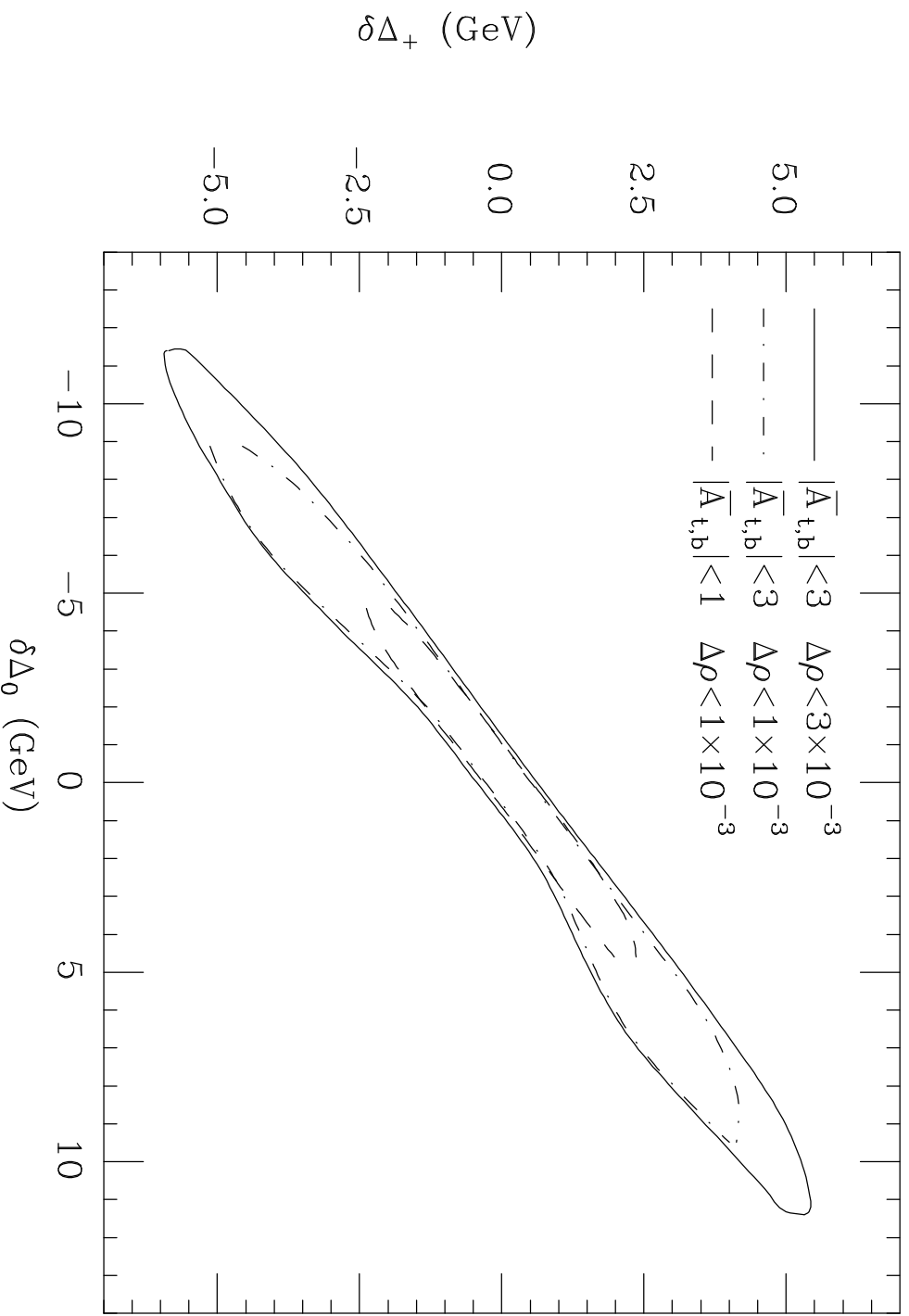
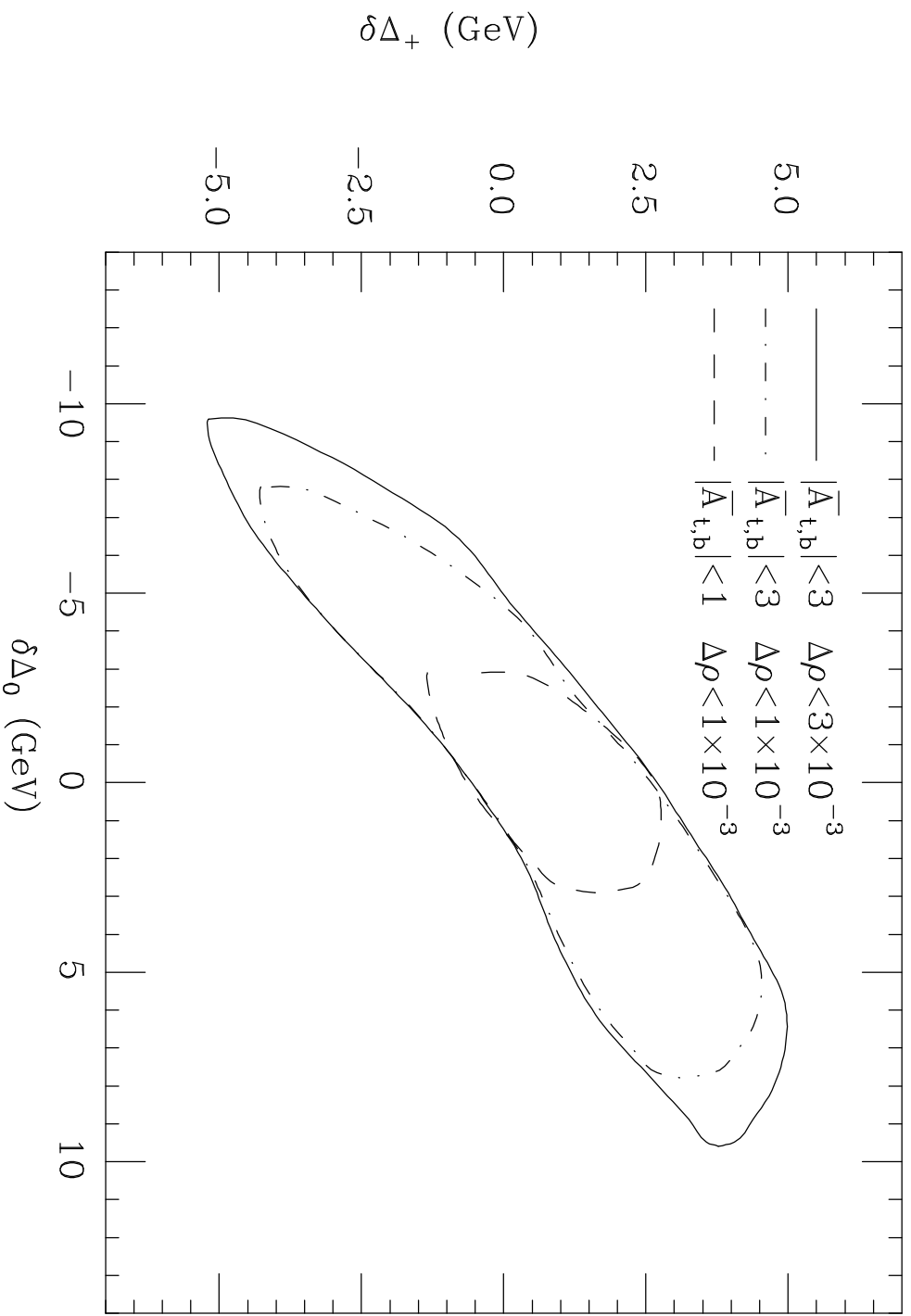
\end{document}